\documentclass[review]{elsarticle}
\usepackage{amsmath, amsfonts, amssymb}
\usepackage[utf8]{inputenc}
\usepackage{graphicx}
\usepackage{geometry}
\usepackage{subfig}
\usepackage{hyperref}
\usepackage{physics}
\usepackage{float}
\usepackage{caption}
\usepackage{cleveref}

\hypersetup{
	bookmarksopen=true,
	colorlinks=true,
	linkcolor=blue,
	citecolor=blue,
	urlcolor=black,	
	pdftitle={Waves in Cosserat media},
	pdfauthor={N. Guarin-Zapata, J. Gomez, C. Valencia, G. Dargush, A.R.  
	Hadjesfandiari},
	pdfkeywords={Finite element method, Cosserat solids, Bloch theorem, Wave propagation},
	pdfsubject={Elastodynamics},
	pdfpagemode=UseOutlines,
	pdfstartview=FitH
}

\journal{Wave motion}
\title{\textbf{Finite element modeling of micropolar-based phononic crystals}}

\author[1]{Nicolás Guarín-Zapata*}
\author[1]{Juan Gomez}
\author[1]{Camilo Valencia}
\author[2]{Gary F. Dargush}
\author[2]{Ali Reza Hadjesfandiari}

\address[1]{Universidad EAFIT, Departamento de Ingeniería Civil, Medellín, 
    Colombia \\* Corresponding author: nguarinz@eafit.edu.co}
\address[2]{Department of Mechanical and Aerospace Engineering, University at 
    Buffalo}

\begin{document}
\begin{frontmatter}

\begin{abstract}
The performance of a Cosserat/micropolar solid as a numerical vehicle to 
represent dispersive media is explored. The study is conducted using the finite 
element method with emphasis on Hermiticity, positive definiteness, principle 
of virtual work and Bloch-Floquet boundary conditions. The periodic boundary 
conditions are given for both translational and rotational degrees of freedom 
and for the associated force- and couple-traction vectors. Results in terms of 
band structures for different  material cells and 
mechanical parameters are provided.
\end{abstract}


\end{frontmatter}

\section*{Introduction}
The increasing growth and strong development of the field of architectured 
materials during recent years has created a renewed interest in generalized or 
extended versions of classical continuum mechanics theories. Popular examples 
of these emerging disciplines can be identified in the area of phononic 
crystals and metamaterials \citep{hussein2014dynamics, srivastava2015elastic}. 
These are materials which by virtue of their architectured microstructure 
exhibit unexpected mechanical properties at the macro level, such as negative 
refraction, negative bulk modulus or negative mass 
\citep{banerjee2011introduction}. For example, the conversion from axial 
deformation into twist, would require chirality, which in turn would require an 
asymmetric stress tensor \cite{lakes2001, frenzel2017}. From the wave 
propagation perspective, these  materials are attractive since they exhibit 
dispersive phenomena, such as filtering and directional effects over fixed 
frequency ranges \citep{gonella2008}. Physically, such phenomena result from 
the interactions of the incident field with the microstructural elements 
producing local scattering and diffraction. Another wave phenomenon of interest 
is cloaking, where the propagation is directed around an object rendering it 
\emph{invisible}. This is another application where there is a need for 
asymmetric stresses. For electromagnetic waves, this can be achieved through 
transformation optics \cite{schurig2006}.


The validation and effective use of these materials, particularly in dynamic 
analysis simulations involves two general steps. In the first place, it is 
necessary to conduct a design or characterization of the material in terms of 
its band structure or frequency-wave vector relation for a representative unit 
cell. This analysis step is typically conducted via Bloch analysis of the unit 
cell, which comprises solving a series of eigenvalue problems for a numerical 
model of the cell that explicitly includes all its microstructural elements. 
The solution from these analyses identify propagation frequencies associated to 
variations of the wave vector along the boundaries of the unit cell after 
considering the spatial periodicity of the material. The second analysis step 
involves the solution of a time-domain boundary value problem, which requires 
the consideration of a large number of material cells filling out the 
particular domain. Clearly, from a numerical point of view, the inclusion of 
the microstructural details in the numerical model implies prohibitive 
computational costs, thus requiring continuum based mechanical models with 
intrinsic capabilities to reproduce dispersive behavior.

Dispersion at the macroscopic level can be understood in terms of scattering 
and diffraction arising when the wave length of the free wave field approaches 
characteristic dimensions of the microstructural elements and, as a result, 
its treatment in terms of continuum mechanics requires theories involving 
constitutive length scale parameters. Broadly speaking, these class of models 
can be classified into (i) gradient based theories and (ii) enriched kinematics 
models. In the former, local kinematic descriptions retaining higher order 
displacement gradients are introduced resulting also in higher order stress 
tensors \citep{TruesdellAndToupin1960, AeroandKuvshinskii1961, Toupin1962, 
MindlinAndTiersten1962, Koiter1964, Mindlin1964, Mindlin1965, 
MindlinAndEshel1965, Eringen1966, yang2002couple, hadjesfandiari2011couple, 
hadjesfandiari2015}. Alternatively, in the latter approach, the material point 
is endowed with additional degrees of  freedom \citep{Cosserat1909, Voigt1910, 
Eringen1966, book:nowacki1986}.

This family of non-classical or generalized continuum models have been used in 
a wide range of applications in several research problems  
\citep{nix1998indentation, fleck1997strain, fleck1994strain, 
stolken1998microbend, Takeo1998,Takeo1998b,TakeoAndIto1997, Midya2004, 
merkel2011, trovalusci2015, iliopoulos2016wave, iliopoulos2017concrete}. 
However, there is a need to examine the strengths and weaknesses of the various 
models and to propose physical experiments that would be helpful in their 
critical assessment.

In this work we explore the capabilities of the micropolar model to capture 
dispersive behavior by virtue of its kinematic variables and particularly 
through the additional constitutive parameters. For that purpose, we address 
several theoretical and simulation aspects relevant to wave propagation in such 
micropolar medium. We start by reviewing the field equations for the model 
with special emphasis placed in the displacement-based equations of motion as 
these reveal phase velocities associated to the possible free wave modes. 
Following that section, we also discuss Bloch periodicity in the context of the 
micropolar theory. We show the relation between displacements and rotations, 
together with its corresponding traction components along the different regions 
of the unit cell properly accounting for the infinite character of the analysis 
domain in terms of a single material cell. Also, as will be shown later, the 
dispersion analysis via the Bloch theorem involves the solution of the 
frequency-domain reduced wave equation and thus it is useful to show the 
Hermitic and positive definite character of the boundary value problem. We 
show these two properties of the operator in section 2.1. The theoretical 
aspects of the paper also describe the variational statement and details for 
its finite element discretization, when conducting Bloch analysis. In the final 
part of the paper, we use the finite element formulation to test the capability 
of the micropolar model to capture dispersive behavior. First, and as a 
verification exercise of the formulation, we find the numerical and analytic 
band structure for a homogeneous micropolar continuum. The homogeneous 
cell analysis is also used to identify appropriate mesh properties in Bloch 
analysis of micropolar media. In a subsequent analysis aimed at producing 
further dispersion in the model, we also consider material cells with simple 
microstructures, namely a bilayer composite and a porous material composed of 
a circular cavity embedded in a micropolar matrix. These are simple 
microstructures, which facilitate the study of the variation of the dispersive 
properties by introducing changes in material and geometric parameters. In both 
cases, we find band structures for different values of the mechanical and 
geometric parameters and track the variation of the cut-off frequency 
associated to the microrotational waves.

\section{Micropolar Model}
The micropolar model used in this work introduces rotational mechanical 
interaction between material points in terms of a couple-tractions vector 
$m_i^{(\hat{n})}$ defined through a generalized Cauchy's postulate as 
\cite{book:mase2009}:
\begin{equation}\label{eq:couple_vector}
	\lim_{\Delta S(\hat n)\to 0} \frac{\Delta M_i}{\Delta S(\hat n)} = m_i^{(\hat n)}
\end{equation}
where $\Delta S(\hat{n})$ is a small element of area oriented with $\hat{n}$ and
$\Delta M_i$ is the resultant moment. The couple-tractions are completely 
described by the couple-stress tensor $\mu _{ij}$ according to
\begin{equation}\label{eq:couple_tractions}
	m_j^{(\hat n)} = \mu_{ij} \hat{n}_i.
\end{equation}

In \eqref{eq:couple_vector} above, $\Delta S(\hat n)$ represents a surface 
material element with outward normal $\hat{n}$. Considering now the classical 
force per unit surface tractions vector $t_j^{(\hat n)}$ related to the Cauchy 
stress tensor $\sigma_{ij}$, such as
\begin{equation}\label{eq:Force_tractions}
	t_j^{(\hat n)} = \sigma_{ij} \hat{n}_i
\end{equation}
leads to momentum and moment of momentum balance equations for the micropolar 
solid \citep{book:nowacki1986}:
\begin{subequations}\label{eq:conservation}
	\begin{align}
	&\sigma_{ji, j} + f_i = \rho \ddot{u}_i\\
	&\sigma_{jk} \epsilon_{ijk} + \mu_{ji, j} + c_i = J \ddot{\theta}_i\, 
	\end{align}
\end{subequations}
and where $ f_i $ and $c_i$ are forces and moments per unit volume; $\rho$ and 
$J$ are the mass and rotational inertial densities, respectively, and 
$\epsilon_{ijk}$ is the Levi-Civita permutation tensor. In the model proposed 
by \cite{hadjesfandiari2011couple} the term $ c_i$ is shown to be equivalent to 
a body force, while $ J $ is eliminated at the onset. Here we retain the 
original form of the equations given in \cite{book:nowacki1986}, where both 
terms are retained. Denoting displacements and microrotation vectors at a field 
point $\vb{x}$ and at the time instant $t$ by $u_i(\vb{x},t)$ and 
$\theta_i(\vb{x},t)$, respectively, we have that the local deformation 
$\gamma_{ji}$ at the material point is now the difference between the 
displacement gradients $u_{i,j}$ and the microrotation (vector) $\theta_k$
\begin{equation}\label{eq:strain}
	\gamma_{ji} = u_{i,j} - \epsilon_{kji} \theta_k\, .
\end{equation}

Also, notice that the consideration of the independent microrotational field 
introduces an additional kinematic variable in the form of a generalized curvature-twist $\kappa 
_{ji}$ describing the change of microrotation per unit length
\begin{equation}\label{eq:curvature}
    {\kappa _{ji}} = {\theta _{i,j}}.
\end{equation}

In a linear isotropic elastic micropolar medium the constitutive equations take 
the following form \citep{book:nowacki1986}:
\begin{subequations}\label{eq:constitutive}
	\begin{align}
	& \sigma_{ji} = (\mu + \alpha) \gamma_{ji} + (\mu - \alpha) \gamma_{ij} + 
	\lambda \gamma_{kk} \delta_{ij}\, ,\\
	& \mu_{ji} = (\eta + \varepsilon) \kappa_{ji} + (\eta - \varepsilon) 
	\kappa_{ij} + \beta \kappa_{kk} \delta_{ij}\, .
	\end{align}
\end{subequations}
where $\mu$ and $\lambda$ are the known Lamé parameters from classical
elasticity, while $\alpha$, $\beta$, $\eta$ and $\varepsilon$ are extra
material parameters from the micropolar model and representative of additional
particle interactions. The set of constitutive equations can also be written in
the following alternative form:
\begin{align*}
& \sigma_{ji} = \mu \gamma^S_{ij} + 2 \alpha \gamma^A_{ij} + \lambda 
\gamma_{kk} \delta_{ij}\, ,\\
& \mu_{ji} = \eta \kappa^S_{ij} + 2\varepsilon \kappa^A_{ij} + \beta 
\kappa_{kk} \delta_{ij}\, ,
\end{align*}
where the superscripts $S$ and $A$ denote the symmetric and skew-symmetric 
parts of the associated second order tensors. We can mention that $\beta$ 
is a parameter only related to torsion, using a parallel with classical 
elasticity it resembles the role of $\lambda$. On the other hand, $\eta$ is 
related to torsion and bending while $\varepsilon$ is only related to bending 
modulus. Furthermore, $\alpha$ is known as micropolar couple modulus and 
quantifies the coupling between micro and macrorotation --- see 
\cite{lakes1991, hassanpour2017, hassanpour2014} for further discussion on the 
interpretation of these parameters.

Using \eqref{eq:constitutive} together with \eqref{eq:strain} and 
\eqref{eq:curvature} in the linear and angular  momentum balance equations 
leads to displacement time-domain equations of motion:
\begin{subequations}
	\begin{align}
	(\lambda + 2\mu) u_{k, ki} - \epsilon_{ijk} \epsilon_{klm} (\mu + \alpha) u_{m,lj}+ 2\alpha \epsilon_{ijk} \theta_{k,j} + f_i = \rho \ddot{u}_i\,  ,\\
	(\beta + 2\eta) \theta_{k, ki} - \epsilon_{ijk} \epsilon_{klm} (\eta + 
	\varepsilon) \theta_{m,lj} + 2\alpha \epsilon_{ijk} u_{k,j} - 
	4\alpha\theta_i + c_i = J \ddot{\theta}_i\, . \label{eq:displa_b}
	\end{align}
	\label{eq:displa}%
\end{subequations}
With boldface characters denoting vector fields, the equations of motion can also be written for completeness in explicit form as:
\begin{subequations}\label{eq:disp_vector}
	\begin{align}
	& (\lambda + 2\mu) \grad\div\vb{u} - (\mu + \alpha)\curl\curl\vb{u} + 
	2\alpha \curl\vb*{\theta} + \vb{f} = \rho \pdv[2]{\vb{u}}{t}, \\
	& (\beta + 2\eta) \grad\div\vb*{\theta} - (\eta + \varepsilon) 
	\curl\curl\vb*{\theta} + 2\alpha \curl\vb{u} - 4\alpha\vb*{\theta} + 
	\vb{c} = J \pdv[2]{\vb*{\theta}}{t}\, . \label{eq:disp_vector_b}
	\end{align}
\end{subequations}
When considering plane problems the term $\grad\div\vb*{\theta}$ is zero and 
the behavior will not depend on the parameter $\beta$, see \ref{app:2Deqs} for 
the explicit expressions for plane problems.

Notice in \eqref{eq:displa_b} and \eqref{eq:disp_vector_b} the appearance of the
dilatation \(\div \vb*{\theta}\), which implicitly assumes that 
\(\vb*{\theta}\) is not, in fact, a pure rotation (or microrotation). On the
other hand, if \(\div \vb*{\theta}=0\) is enforced, then the resulting theory
would suffer from the same type of indeterminacy as the original couple stress
theory \citep{MindlinAndTiersten1962}. Following the current convention, 
however, \(\vb*{\theta}\) will continue to be called the microrotation in this
paper.

In order to conduct Bloch analysis, and more specifically to determine the 
dispersion relations for a micropolar solid, it is convenient to neglect the 
body force and couple densities and to assume a time dependence of the form 
$e^{-i\omega t}$ for both the displacement and rotation field, therefore 
yielding the following reduced frequency-domain version of the equations of 
motion:
\begin{equation}
\begin{aligned}
  (\lambda + 2\mu)\grad\div\vb{u}- (\mu + \alpha)\curl\curl\vb{u} + 2\alpha\curl\vb*{\theta} &= -\rho\omega^2 \vb{u} \, ,\\
  (\beta + 2\eta)\grad\div\vb*{\theta} - (\eta + 
  \varepsilon)\curl\curl\vb*{\theta} + 2\alpha\curl\vb{u} - 4\alpha\vb*{\theta} 
  &= - J \omega^2 \vb*{\theta} \, .
\end{aligned}
\label{eq:wave_eq_freq}
\end{equation}
For convenience in later developments and to aid the comparison with similar 
formulations available in the literature, it is convenient to use also the 
alternative form:
\begin{equation}
\begin{aligned}
c_1^2\grad\div\vb{u}- c_2^2\curl\curl\vb{u} + K^2\curl\vb*{\theta} &= -\omega^2 \vb{u} \, ,\\
c_3^2\grad\div\vb*{\theta} - c_4^2\curl\curl\vb*{\theta} + Q^2\curl\vb{u} - 2Q^2\vb*{\theta} &= -\omega^2 \vb*{\theta} \, 
\end{aligned}
\label{eq:wave_eq_speeds}
\end{equation}
and where: $c_1$ represents the phase/group speed for the longitudinal wave 
($P$) that is non-dispersive as in the classical case, $c_2$ represents the 
high-frequency limit ($k\rightarrow\infty$) phase/group speed for a transverse 
wave ($S$) that is dispersive unlike the classical counterpart, $c_3$ 
represents the high-frequency limit ($k\rightarrow\infty$) phase/group speed 
for a longitudinal-rotational wave ($LR$) with a corkscrew-like motion that is 
dispersive and does not have a classical counterpart, $c_4$ represents 
the high-frequency limit ($k\rightarrow\infty$) phase/group speed for a 
transverse-rotational wave ($TR$) that is dispersive and does not have a 
classical counterpart, $Q$ represents the cut-off frequency for rotational 
waves appearance, and $K$ quantifies the difference between the low-frequency 
and high-frequency phase/group speed for the S-wave ---see 
\cref{fig:dispersion_qual} for a qualitative description. These parameters are 
defined by:
\begin{equation*}
\begin{split}
c_1^2 = \frac{\lambda +2\mu}{\rho},\quad &c_3^2 =\frac{\beta + 2\eta}{J},\\
c_2^2 = \frac{\mu +\alpha}{\rho},\quad &c_4^2 =\frac{\eta + \varepsilon}{J},\\
Q^2= \frac{2\alpha}{J},\quad &K^2 =\frac{2\alpha}{\rho} \, .
\end{split}
\end{equation*}
We should highlight that $c_3$ does not play a role for waves in the plane. As 
mentioned before, $\grad\div\vb*{\theta} = 0$ in that case.

The Principle of Virtual Work (PVW) for a micropolar solid follows after 
considering the translational and rotational equilibrium equations, using the 
virtual fields $\delta u_i$ and $\delta \theta_i$ as weighting functions and 
integrating over the volume $V$ with the aid of the divergence theorem. 
Denoting the kinematic measures in a micropolar solid by $\gamma^S_{ij}$, 
$\gamma_{ij}^{A}$ and $\kappa_{ij}$ corresponding to the classical 
infinitesimal strain tensor, the skew-symmetric part of the relative 
deformation tensor --- i.e., the difference between the displacement gradient 
and the micro-displacement gradient --- and the generalized curvature-twist 
tensor allow us to write this  principle in the form
\begin{equation}
\begin{split}
\int\limits_{V} \sigma^S_{ij} \delta\gamma^S_{ij}\dd{V} + \int\limits_{V} 
\sigma^A_{ij}\delta\gamma_{ij}^{A} \dd{V} + 
\int\limits_{V}\mu_{ij}\delta\kappa_{ij}\dd{V} - \int\limits_{S} t_i\delta_i 
\dd{S} 
- \int\limits_{S}m_i \delta\theta_i \dd{S} -\\
 \omega^2\int\limits_{V} \rho u_i \delta u_i \dd{V} - 
 \omega^2\int\limits_{V} J \theta_i\delta\theta_i \dd{V} = 0
\end{split}
\label{eq:PVW}
\end{equation}
where $\sigma^S_{ij}$ is the symmetric part of the stress tensor, while
$\sigma^A_{ij}$ is the skew-symmetric part of the stress tensor.

\section{Formulation for Periodic Materials}
Here we review the relevant aspects of the analysis of spatially periodic 
materials in terms of the theory of phononic crystals, and particularly the 
so-called Bloch-Floquet periodic boundary conditions applied to the micropolar 
solid. For an in-depth discussion of periodic materials the reader is referred 
to classical sources \citep{book:brillouin, book:kittel}, while a comprehensive 
review is provided in \cite{hussein2014dynamics}. In that theory, the key 
concept is established by Bloch's theorem, providing a relationship between the 
fields on opposite sides of the cell and taking into account spatial 
periodicity in a wave propagation problem. Therefore, the characterization of 
the micropolar medium is to be conducted after assuming that the material is 
the result of the spatial and periodic repetition of a fundamental unit cell. 
Under this assumption, a fundamental cell containing a motif repeats itself (in 
one, two, or three space dimensions) according to a spatial period defined in 
terms of a lattice vector. The \emph{motif} refers to a microstructural 
heterogeneity which could contain different materials and geometries, as well 
as 
fluids and/or solids of the classical or Cosserat type. The dispersive 
properties of such a periodic material, given in terms of frequency-wave number 
relations (or band diagram), can be found from the analysis of a single 
fundamental cell after using Bloch's theorem, which establishes that a function 
$ \mathbf{u}(\mathbf{x})$ can be expressed in the form
\begin{equation}
  \mathbf{u}(\mathbf{x}) = \mathbf{w}(\mathbf{x}) e^{i\mathbf{k}\cdot \mathbf{x}}\, ,
  \label{eq:bloch}
\end{equation}
where $\mathbf{w}(\mathbf{x})$ is the  Bloch function that has the same 
periodicity of the material and $\mathbf{k}$ is a wave vector. Accordingly, the 
solution is the product of a periodic function with the periodicity of the 
lattice and a plane wave, that is also periodic. As a consequence, the field 
variables in the differential equation satisfy the relation
\[\mathbf{\Phi}(\mathbf{x} + \mathbf{a}) = \mathbf{\Phi}(\mathbf{x})e^{i\mathbf{k}\cdot\mathbf{a}}\, ,\]
connecting the variable $\mathbf{\Phi}$ at opposite sides of the unit cell set 
apart by a vector $\mathbf{a}$. In this case, $\mathbf{\Phi}$ refers to the 
principal variable (or any of its spatial derivatives) involved in the physical 
problem. It then follows that if one wants to characterize the material in 
terms of its wave propagation velocities to obtain values that can be used in a 
homogenized continuum model, it  suffices to  analyze a single cell. In the 
case of the micropolar medium, Bloch's theorem states that the eigenfunctions 
of (\ref{eq:wave_eq_freq}) can be expressed in the form
\begin{align*}
&u_r(\vb{x}) = u_r(\vb{x}+\vb{a})e^{i\vb{k}\cdot\vb{a}}, \\ 
&\theta_r(\vb{x}) = \theta_r(\vb{x}+\vb{a})e^{i\vb{k}\cdot\vb{a}},
\end{align*}
where $\vb{a}$ is a vector that represents the periodicity of the material. 
That is, the solution is the same at opposite sides of the unit cell, except 
for a phase shift factor $e^{i\vb{k}\cdot\vb{a}}$. Due to the linearity of the 
differential equations we also have the following Bloch-periodic boundary 
conditions for the corresponding traction vectors
\begin{align*}
&t_r(\vb{x}) = -t_r(\vb{x}+\vb{a})e^{i\vb{k}\cdot\vb{a}}, \\ 
&m_r(\vb{x}) = -m_r(\vb{x}+\vb{a})e^{i\vb{k}\cdot\vb{a}} .
\end{align*}

Thus, in the case of the micropolar solid, Bloch's theorem reduces to the 
following set of boundary conditions for displacements, microrotations, 
force-tractions and couple-tractions:
\begin{subequations}
\begin{align}
&u_r(\vb{x}) = u_r(\vb{x}+\vb{a})e^{i\vb{k}\cdot\vb{a}}\, , \\ 
&\theta_r(\vb{x}) = \theta_r(\vb{x}+\vb{a})e^{i\vb{k}\cdot\vb{a}}\, ,\\
&t_r(\vb{x}) = -t_r(\vb{x}+\vb{a})e^{i\vb{k}\cdot\vb{a}}\, , \\ 
&m_r(\vb{x}) = -m_r(\vb{x}+\vb{a})e^{i\vb{k}\cdot\vb{a}}\, .
\end{align}
\label{eq:bloch_bcs}
\end{subequations}

The set of conditions summarized in \eqref{eq:bloch_bcs} will be satisfied in a 
variational sense using a finite element formulation. Subsequently, a numerical 
model of the unit cell resulting in a generalized eigenvalue problem will be 
solved for various specifications of the wave vector. The details of such an
implementation will be discussed next.

\subsection{Hermiticity of the equations}
Since the dynamic analysis of the micropolar medium involves solution of the 
frequency domain reduced wave equation subject to Bloch periodic boundary 
conditions, as given by \eqref{eq:bloch_bcs}, it becomes necessary to rewrite 
the PVW \eqref{eq:PVW} so it can be used to properly represent inner products 
in complex-valued vector spaces. Using the operator $*$ to denote complex 
conjugate, we can write:
\begin{equation}
\begin{split}
\delta\Pi(\omega,u_i,\hat{u}_i)=\int\limits_{V} 
\tau_{ij}^{*}\hat\epsilon_{ij}\dd{V} + \int\limits_{V} 
\sigma_{ij}^{*}\hat\gamma_{ij}^{A}\dd{V} + 
\int\limits_{V}\mu_{ij}^{*}\hat\kappa_{ij}\dd{V} - \int\limits_{S} 
t_i^{*}\hat{u}_i \dd{S} \\
- \int\limits_{S}m_i^{*} \hat\theta_i \dd{S} - 
\omega^2\int\limits_{V} \rho u_i^{*} \hat{u}_i \dd{V} - 
\omega^2\int\limits_{V} J\theta_i^{*}\hat\theta_i \dd{V} \equiv 0 \, .
\end{split}
\label{eq:MPVW}
\end{equation}

In the principle of virtual work given by \eqref{eq:MPVW} $u_i$ and $\hat{u}_i$ 
represent the actual and virtual fields respectively. In the sense of 
\eqref{eq:MPVW}, $\delta\Pi$ could be understood as the variation of the 
Lagrangian functional of the micropolar system. As will be demonstrated next, 
if we interchange the variables $u_i$ and $\hat{u}_i$, we find that
\[ \delta\Pi(\omega,u_i,\hat{u}_i) = \delta\Pi(\omega,\hat{u}_i,u_i) \,\] which implies that the operator is Hermitian (self-adjoint) under Bloch periodic boundary conditions also resulting in Hermitian matrices when discretized via finite elements.

The proof of Hermiticity follows after one uses Bloch periodicity conditions 
between the tractions and displacements (see \eqref{eq:bloch_bcs}) into the 
boundary terms in \eqref{eq:MPVW}, which yields
\begin{equation}
\begin{split}
&\int\limits_S t_r^{*}(\vb{x})u_r(\vb{x}) \dd{S} + \int\limits_S 
m_r^{*}(\vb{x})\theta_r(\vb{x}) \dd{S} = \\
&\sum\limits_p \left\lbrace \int\limits_{S_p}\left[t_r^{*}(\vb{x})u_r(\vb{x}) + 
t_r^{*}(\vb{x}+\vb{a}_p)u_r(\vb{x}+\vb{a}_p)\right]\dd{S}_p + \right.\\
&\left. \int\limits_{S_p}\left[m_r^{*}(\vb{x})\theta_r(\vb{x}) + 
m_r^{*}(\vb{x}+\vb{a}_p)\theta_r(\vb{x}+\vb{a}_p)\right]\dd{S}_p \right\rbrace 
\, ,
\end{split}
\end{equation}
where the index $p$ refers to each pair of opposite sides of the boundary. 
Introducing the phase shifts and elaborating further gives:
\begin{equation}
\begin{split}
&\int\limits_S t_r^{*}(\vb{x})u_r(\vb{x}) \dd{S} + \int\limits_S 
m_r^{*}(\vb{x})\theta_r(\vb{x}) \dd{S} = \\
&\sum\limits_p \left\lbrace \int\limits_{S_p} u_r(\vb{x})\left[t_r^{*}(\vb{x}) 
+ e^{-i\vb{k}\cdot\vb{a}}t_r^{*}(\vb{x}+\vb{a}_p)\right]\dd{S}_p + \right.\\
&\left. \int\limits_{S_p}\theta_r(\vb{x})\left[m_r^{*}(\vb{x}) + 
e^{-i\vb{k}\cdot\vb{a}}m_r^{*}(\vb{x}+\vb{a}_p)\right]\dd{S}_p \right\rbrace \,
\end{split}
\end{equation}
which after taking the complex conjugate reduces to the Bloch-equilibrium 
condition (or the relationship between the traction vectors at opposite faces 
of the cell) for the terms enclosed by the square brackets. This leads to the
vanishing of the boundary terms proving the Hermiticity condition.

\subsection{Positive definiteness}
To demonstrate positive (semi)-definiteness it is convenient to define the 
total potential and kinetic energy functionals $U([u,\theta],[u,\theta])$ and 
$T([u,\theta],[u,\theta])$ such that
\[\begin{split}
U([u,\theta],[u,\theta]) &= \int\limits_{V} \gamma_{ij}^{S*} 
C_{ijkl}\gamma^S_{ij}\dd{V} + \int\limits_{V} \alpha 
\gamma_{ij}^{A*}\gamma_{ij}^{A}\dd{V} + 
\int\limits_{V}\kappa_{ij}^{*}D_{ijkl}\kappa_{ij}\dd{V} \, ,\\
T([u,\theta],[u,\theta]) &= \int\limits_{V} \rho u_i^{*} u_i \dd{V} + 
\int\limits_{V} J \theta_i^{*}\theta_i \dd{V} \, .
\end{split}\]

For the potential functional to be positive definite we need that the 
constitutive tensors are positive definite. For an isotropic material, this
implies the following constraints in the material parameters:
\begin{align*}
  &\mu >0\, ,\quad  \alpha > 0\, ,\quad \eta >0\, ,\quad \varepsilon >0\, ,\\
  &3\lambda + 2\mu >0\, ,\quad 3\beta + 2\eta >0\,  . 
\end{align*}

It should be noted that there exist differences in notation in the literature
for material parameters. Particularly, the symbols used by \cite{Eringen1966}
are known to be confusing since the symbol \(\mu\) is used for
a combination of the classic Lamé parameter and a micropolar parameter. The use
of this symbol have led to incorrect inequalities, as presented by
\cite{cowin1970b} and \cite{hassanpour2017}.

In order to have the condition  $[u,\theta]\neq 0$, the functional $T$ should 
be different from zero implying that
\begin{equation}
\omega_r^2 = 
\frac{U([u_r,\theta_r],[u_r,\theta_r])}{T([u_r,\theta_r],[u_r,\theta_r])} \, ,
\label{eq:definiteness}
\end{equation}
meaning that $\omega$ is always greater than or equal to zero. This result is 
in agreement with the physical meaning of angular frequency given to $\omega$. 
On the other hand, the potential energy could be zero in the case of rigid body 
motion implying that the form $U$ is positive semi-definite, while the form $T$ 
is positive definite and so are their discrete counterparts.

\section{FEM formulation}
Finite element equations for the micropolar solid are straightforward to 
obtain after introducing the displacement and microrotation interpolation 
functions $_{u}\hspace{-1pt}N_{i}^{k}$ and $_{\theta}\hspace{-1pt}N_i^{k}$. 
Here the superscript $k$ makes reference to the contribution from the $k$th 
node of a given element, while the right subscript $i$ indicates the tensorial 
nature of the variable being interpolated. The displacement and microrotation 
vectors at a given point inside the element can now be written in terms of the 
nodal variables $u^k$ and $\theta^k$ as
\begin{equation}
u_i = _{u}\hspace{-4pt}N_{i}^{k}u^{k},\quad \theta_i = _{\theta}\hspace{-4pt}N_i^{k}\theta^{k} \, .
\label{eq:interp}
\end{equation}
where the summation convention for repeated indices applies for both physical 
and interpolation subscripts and superscripts. For the actual numerical 
implementation, it is convenient to express the microrotation vector 
$\theta_i$ in terms of its dual skew-symmetric rotation tensor $\theta_{ij}$, 
such that
\[\theta_{ij} = \epsilon_{qij}\ _{\theta}N_q^k \theta^k \equiv R_{ij}^k \theta^k \, .\]

Kinematic descriptions in terms of derivatives of the primary fields $u_i$ and 
$\theta_i$ follow from suitable combinations of derivatives of the primary 
interpolation functions. Thus,
\begin{equation}
\begin{split}
\gamma^S_{ij} = B_{ij}^k u^k,\quad \omega_{ij} = \hat{W}_{ij}^k u^k,\\
\kappa_{ij} = M_{ij}^k \theta^k,\quad \gamma_{ij}^A = \hat{W}_{ij}^k u^k + R_{ij}^k \theta^k .
\end{split}
\label{eq:diff_interp}
\end{equation}
where the tensor $\omega_{ij}$ denotes the skew-symmetric part of the 
displacement gradient. Substitution of (\ref{eq:interp}) and 
(\ref{eq:diff_interp}) into (\ref{eq:PVW}) yields the PVW in terms of virtual 
nodal variables $\delta u^k$ and $\delta \theta^k$
\begin{equation}
\begin{split}
\delta u^k \int\limits_{v} B_{ij}^k \sigma^S_{ij} \dd{V} + \delta 
u^k\int\limits_{V} \hat{W}_{ij}^k \sigma^A_{ij} \dd{V} + \delta 
\theta^k\int\limits_{V} R_{ij}^k \sigma^A_{ij} \dd{V} \\
+ \delta\theta^k\int\limits_{V} M_{ij}^k \mu_{ij} \dd{V}
- \omega^2\delta u^k\int\limits_{V} \rho \ _{u}N_{i}^k u_i \dd{V} - \omega^2 
\delta\theta^k \int\limits_{V} J\ _{\theta}N_{i}^{k} \theta_i \dd{V}\\
 - \delta u^k\int\limits_{S}\ _{u}N_{i}^k t_i \dd{S} - \delta 
 \theta^k\int\limits_{S}\ _{\theta}N_i^k m_i \dd{S} = 0 \, .
\end{split}
\label{eq:discrete_PVW}
\end{equation}

Using the arbitrary character of the virtual fundamental fields $\delta u^k$ 
and $\delta \theta^k$ in \eqref{eq:discrete_PVW}, gives the following set of 
weak equilibrium equations in terms of nodal forces and couples consistent with 
the stresses and couple stresses;
\begin{equation}
\begin{aligned}
\hat{f}_{\tau}^k + \hat{f}_{\sigma}^k - \hat{f}_I^k-\hat{T}^k &= 0\, ,\\
\hat{m}_{\sigma}^k + \hat{m}_{\mu}^k-\hat{m}_I^k-\hat{q}^k &= 0\, .
\end{aligned}
\label{eq:FORMOM}
\end{equation}
where the different terms become obvious after comparing 
\eqref{eq:discrete_PVW} and \eqref{eq:FORMOM}. These equations describing 
equilibrium of forces and moments for the $k$-th degree of freedom can be 
written in the following matrix form
\begin{equation}
\left[ \begin{array}{cc}
K_{uu}^{kp} & K_{u\theta}^{kp} \\ 
K_{\theta u}^{kp} & K_{\theta\theta}^{kp}
\end{array} \right] \left\lbrace \begin{array}{c}
u^p \\ 
\theta^p
\end{array} \right\rbrace - \omega^2\left[ \begin{array}{cc}
M_{uu}^{kp} & 0 \\ 
0 & M_{\theta\theta}^{kp}
\end{array}  \right]\left\lbrace \begin{array}{c}
u^p \\ 
\theta^p
\end{array}\right\rbrace = \left\lbrace \begin{array}{c}
f_{ext} \\ 
q_{ext}
\end{array}\right\rbrace \, ,
\end{equation}
which results after writing the stress-strain relationships for the micropolar 
solid in terms of constitutive tensors $C_{ijkl}$, $G_{ijkl}$ and $D_{ijkl}$ as
\begin{equation}
\begin{aligned}
  &\sigma_{ij} = C_{ijkl}\gamma^S_{kl} + G_{ijkl}\gamma_{kl}^A\, , \\
  &\mu_{ij} = D_{ijkl}\kappa_{kl}\, .
\end{aligned}
\label{eq:const}
\end{equation}
For completeness, all of the terms in the matrix equation for the 
micropolar solid are given in Appendix A.

Imposition of the Bloch periodic boundary conditions in the system given by 
\eqref{eq:FORMOM} results in a generalized eigenvalue problem of the form 
\citep{guarin2015}:
\begin{equation}
\left[K_R - \omega^2M_R\right] \{ \textbf{U}_R \} = \textbf{0}\, .
\label{eq:eig}
\end{equation}

\section{Dispersion relations for a micropolar cellular material}
We now conduct a series of numerical simulations intended to test the 
capabilities of the micropolar model as a numerical vehicle to introduce 
dispersive behavior through a continuum based approach. For that purpose, we 
consider first the simplest case of a homogeneous material cell. This ideal 
case is also useful as a verification problem for the numerical implementation 
as that model has a closed-form dispersion relation. At the same time, the 
homogeneous case is used to assess the convergence of the band structure 
predicted by the numerical model. The particularization of the equations of 
motion \eqref{eq:displa} to the in-plane 2D problem is described in 
\ref{app:2Deqs}. In a subsequent analysis, and intended to identify the 
sensitivity of the micropolar material parameters in the band structure, we 
also performed Bloch analysis for a bilayer composite. In this case, the band 
structures were found for different values of a single mechanical parameter, 
while keeping constant values for the remaining ones. As a final test, we 
extended our analysis to a cellular material with a microstructure 
corresponding 
to a circular pore embedded in an otherwise micropolar medium. In this case, we 
wanted to test the sensitivity of the dispersive response to the relative size 
of the pore with respect to the length scale constitutive parameter implicit in 
the material model.

\subsection{Homogeneous material}
\begin{figure}[H]
    \centering
    \includegraphics[height=3 in]{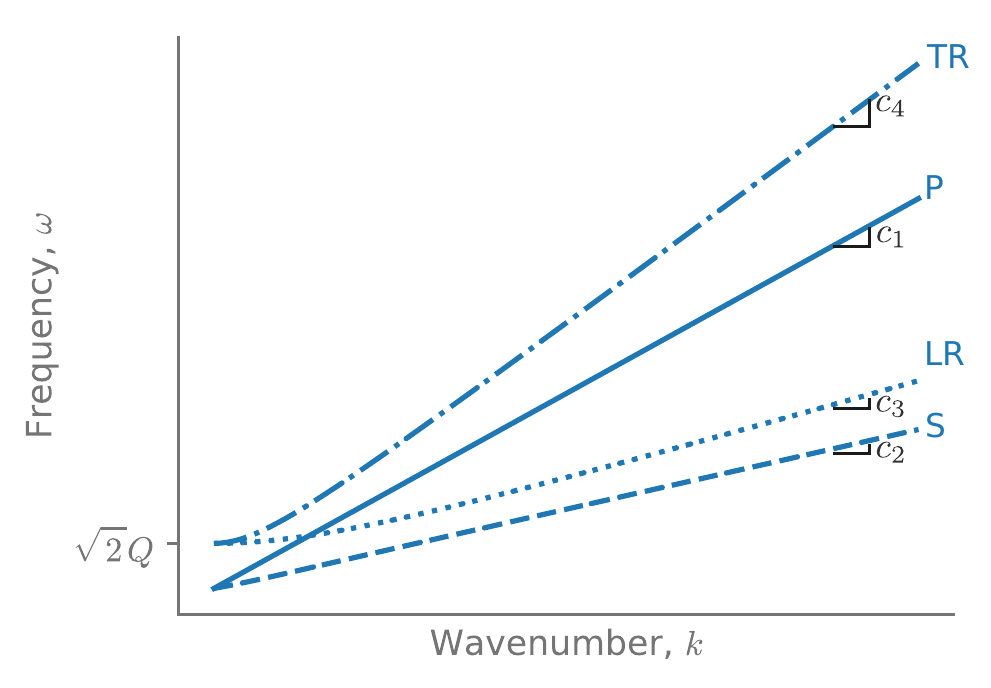}
    \caption{Qualitative depiction of the dispersion relations for a micropolar 
    material. The different branches are: ($P$) non-dispersive 
    longitudinal wave with phase/group speed $c_1$ as in the classical case, 
    ($S$) dispersive transverse wave with limiting ($k\rightarrow\infty$) 
    phase/group speed $c_2$, ($LR$) dispersive longitudinal-rotational 
    (corkscrew-like) wave with limiting ($k\rightarrow\infty$) phase/group 
    speed $c_3$, ($TR$) dispersive transverse-rotational wave wave with 
    limiting ($k\rightarrow\infty$) phase/group speed $c_4$.}
    \label{fig:dispersion_qual}
\end{figure}

In a homogeneous micropolar solid, dispersion relations can be obtained in 
closed-form (See \cref{fig:dispersion_qual}). Following, we present the 
expressions for a two-dimensional solid.
The frequency-wavenumber relationships can be written in compact form as
\begin{align*}
&\omega^P_{m, n} =  c_1 k_{m, n}\, ,\\
&\omega^{S}_{m, n} =  \sqrt{\frac{A}{2} - \frac{1}{2}\sqrt{A^2 - 4B}}\, ,\\
&\omega^{TR}_{m, n} =  \sqrt{\frac{A}{2} + \frac{1}{2}\sqrt{A^2 - 4B}}\, ,
\end{align*}
where the constants $A$ and $B$ correspond to
\begin{align*}
&A = 2Q^2 + (c_2^2 + c_4^2) k_{m, n}^2\, ,\\
&B = 2Q^2 c_2^2 k_{m, n}^2 - K^2 Q^2 k_{m, n}^2 + c_2^2 c_4^2 k_{m, n}^4\, ,
\end{align*}
and TR refers to transverse-rotational.

One difference between wave propagation in micropolar and classical elasticity 
is the appearance of new (rotational) propagating waves that are dispersive. 
These waves appear above a cut-off frequency given by $\omega_0^2 = 2Q^2$ 
\citep{book:nowacki1986}. Besides the two propagation modes mentioned above 
there is another one with the following frequency-wavenumber relation
\[\omega^{LR}_{m,n} = \sqrt{2Q^2 + c_3^2 k_{m, n}^2}\, ,\]
where LR refers to longitudinal-rotational. This wave only exists in 3D, and 
will not be present in the following results.

When conducting Bloch analysis for a 2D cell, the Bloch theorem requires that
the wave number vector is swept over the first Brillouin zone
\citep{book:brillouin}. As a result, the $m$ and $n$ subscripts 
in the frequency and wave number terms represent values of the wave number 
along adjacent Brillouin zones and refer to waves coming from these adjacent 
Brillouin zones. These wave numbers are given by:
\begin{equation}
k_{m,n} = \sqrt{\left(k_x + \frac{m\pi}{d}\right)^2
    + \left(k_y + \frac{n\pi}{d}\right)^2}\, .
\label{eq:wavenumber_adjacent}
\end{equation}

\begin{figure}[H]
    \centering
    \includegraphics[height=1.8 in]{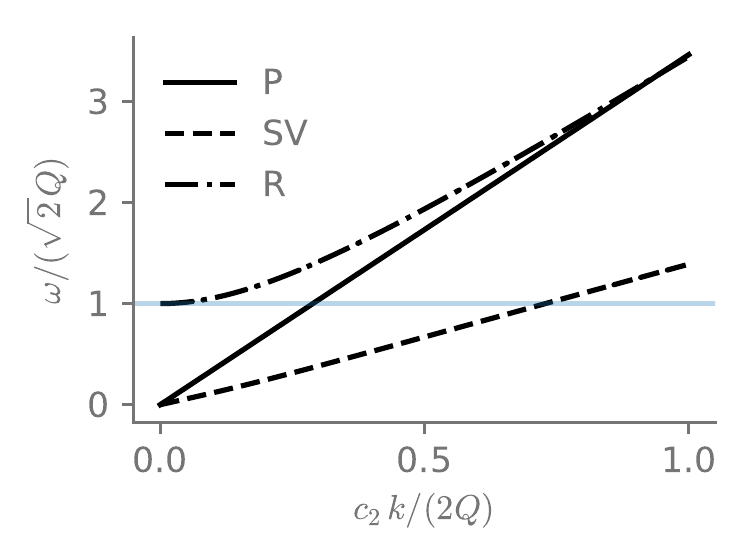}
    \includegraphics[height=1.8 in]{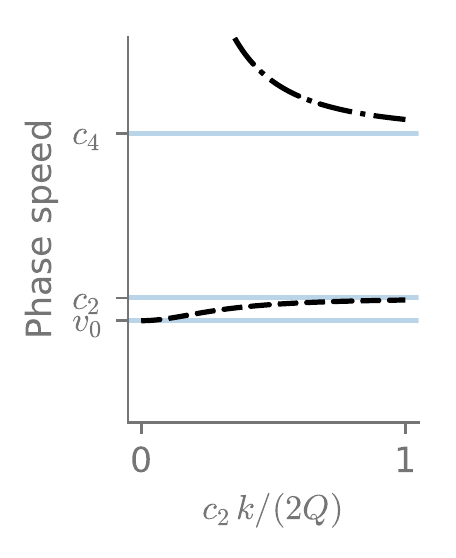}
    \includegraphics[height=1.8 in]{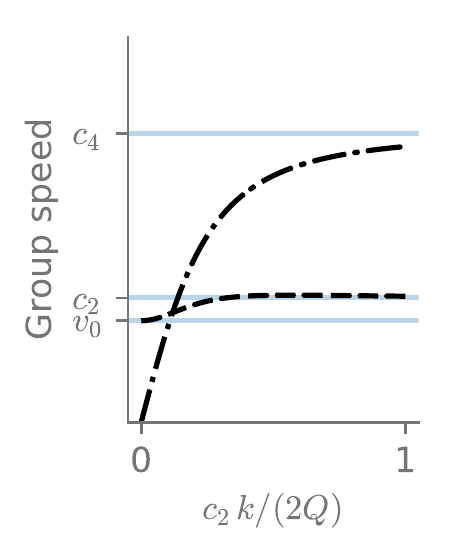}
    \caption{Dispersion relations for a homogeneous micro-polar material 
    material with properties: $\rho = 1\times 10^5$, $J=1\times 10^4$, 
    $\lambda=2.8 \times 10^{10}$, $\eta + \varepsilon=1.62 \times 10^9$, $\mu = 
    4\times 10^9$, $\alpha = 2 \times 10^9$ as in \cite{kulesh2009}. The plot 
    on the left shows the frequency-wave number relation for the non-dispersive 
    P-wave (continuous line) and the dispersive SV and micro-rotational wave 
    (dashed lines). The microrotational wave is only triggered above the 
    normalized frequency of 1. The plots in the middle and right part of the 
    figure show the phase and group speeds for the dispersive modes.}
    \label{fig:kulesh_analytic}
\end{figure}

Typical dispersion relationships are shown in \cref{fig:kulesh_analytic} for the mechanical parameters reported in \cite{kulesh2009} and corresponding to:
\begin{align*}
&\rho = 10^5 \mbox{ kg/m}^3,\quad &\lambda = 2.8\times 10^{10} \mbox{ 
N/m}^2,\qquad &\mu = 4\times 10^9 \mbox{ N/m}^2,\\
&J = 10^4 \mbox{ kg/m},\quad &\eta + \varepsilon = 1.62\times 10^9 \mbox{ 
N},\qquad &\alpha = 2\times 10^9 \mbox{ N/m}^2.
\end{align*}

The figure \ref{fig:kulesh_analytic} shows the cut-off frequency associated 
with the microrotational 
wave, together with the limit cases for the phase and group speeds. 

\begin{figure}[H]
    \centering
    \includegraphics[width=6 in]{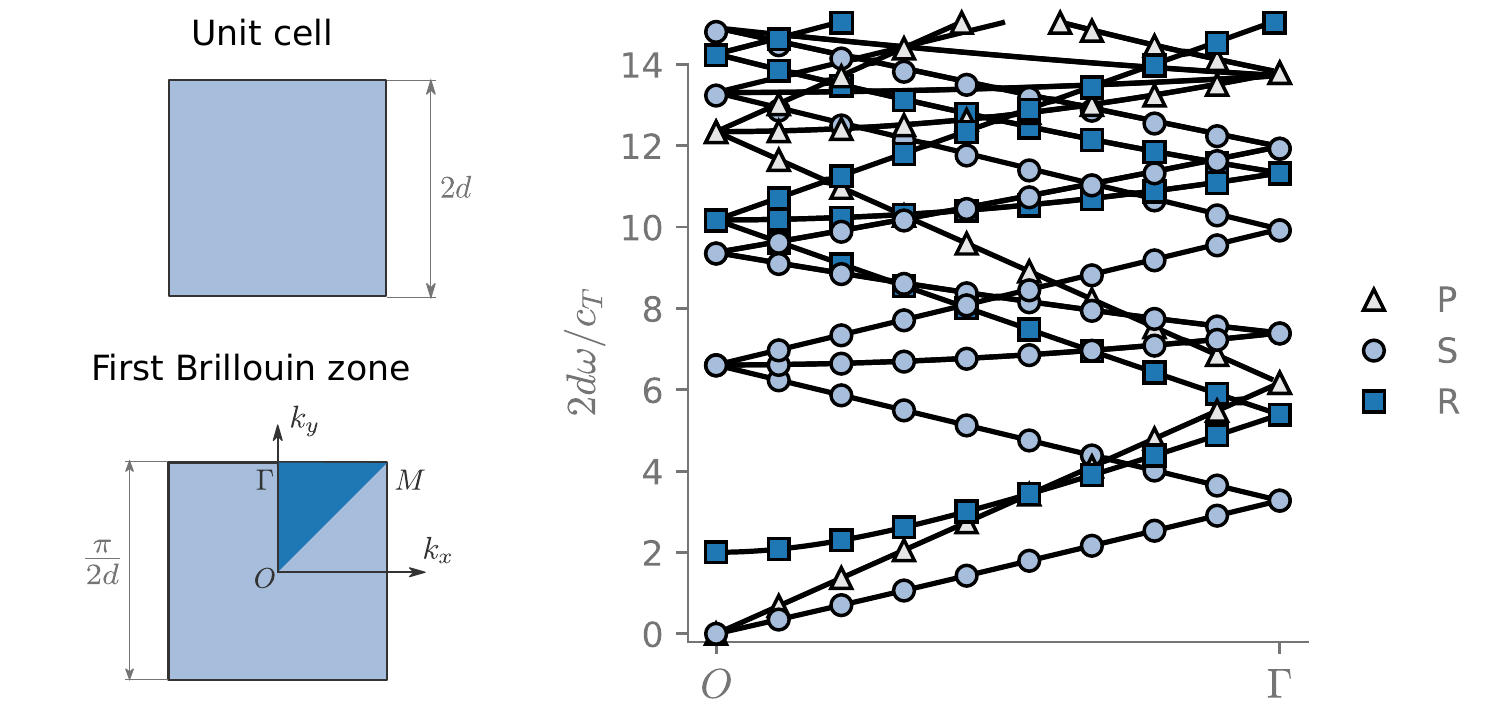}
    \caption{Dispersion relations for a homogeneous micro-polar material model. 
    Solid lines represent FEM results while markers correspond to the analytic 
    solution. The left part of the figure shows the unit material cell and the 
    first Brillouin zone.}
    \label{fig:homogeneous}
\end{figure}

Figure \ref{fig:homogeneous} compares now the analytic and numerical band structure for the micro-polar solid for the following set of material parameters:
\begin{align*}
\rho = 2770 \mbox{ kg/m}^3,\quad &\lambda = 5.12\times 10^{10} \mbox{ N/m}^2,\quad &\mu = 2.76\times 10^{10} \mbox{ N/m}^2,\\
J = 306.5 \mbox{ kg/m},\quad &\eta + \varepsilon = 7.66\times 10^9 \mbox{ 
N},\quad &\alpha = 3.07\times 10^9 \mbox{ N/m}^2\, .
\end{align*}

 The numerical curves were obtained with a mesh of $34\times34$ bilinear 
 elements. The values for the classical model material parameters are those of 
 aluminum, while the ones for the micropolar model have been adjusted to yield 
 a normalized cut-off frequency of 2. The figure also shows the unit material 
 cell and the first Brillouin zone. The numerical implementation accurately 
 predicts the propagation modes, including the micro-rotational wave, together 
 with the analytic value of the cut-off frequency.

\begin{figure}[H]
    \centering
    \includegraphics[width=6 in]{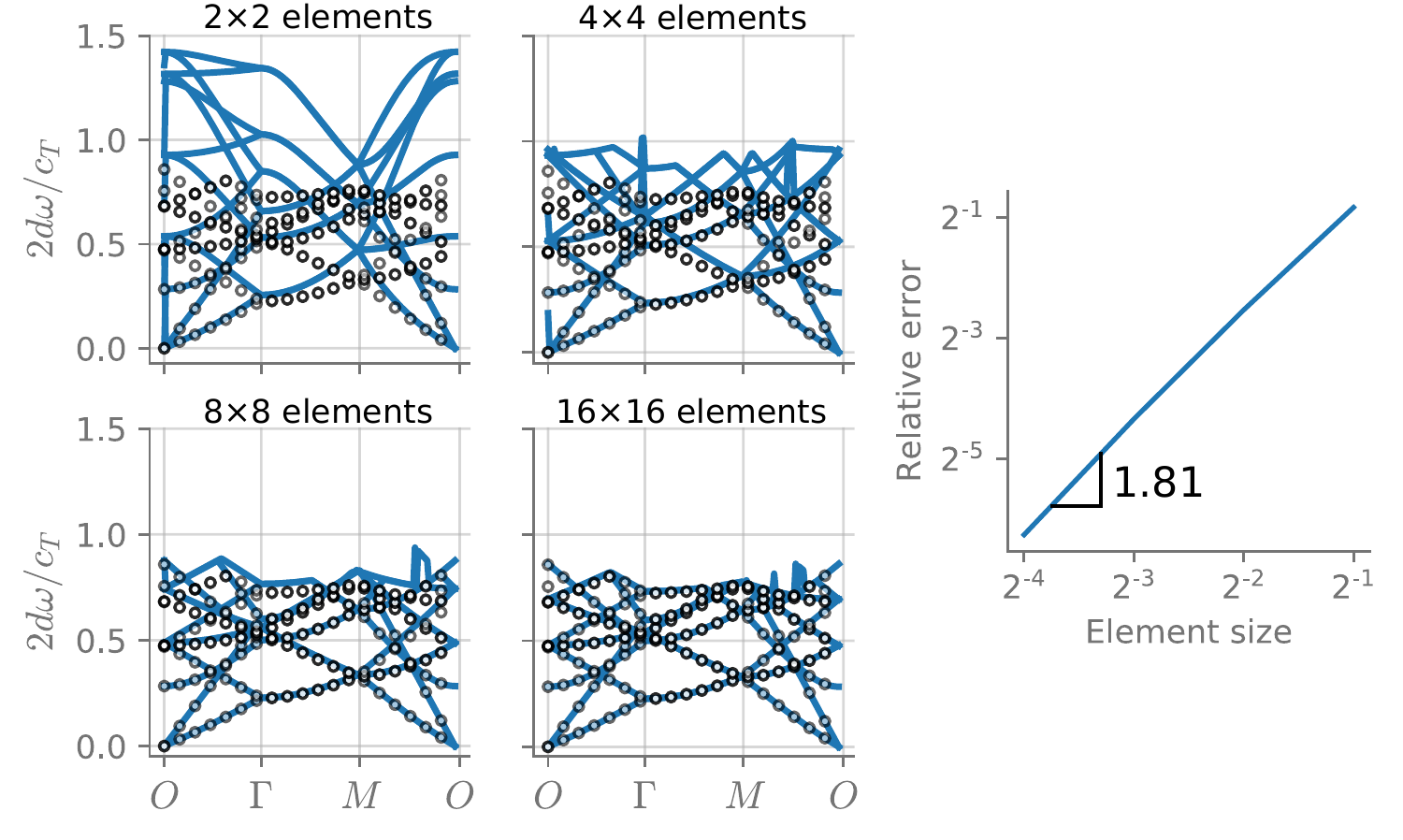}
    \caption{Convergence of the first 12 modes in the dispersion relations for 
    a sequence of meshes with: $2\times2$, $4\times4$, $8\times8$, and 
    $16\times16$ elements --- presented as solid blue lines in the background. 
    The results are compared with a mesh that has $32\times32$ elements --- 
    presented as dots in the foreground.  The estimated convergence rate for 
    the eigenvalues in the 2-norm is 1.81.}
    \label{fig:convergence}
\end{figure}

As an additional verification, we also tested the convergence in the 
calculation of the dispersion relations after considering the first 12 modes 
for a sequence of meshes of $2\times2$, $4\times4$, $8\times8$, and 
$16\times16$ elements. The error in the eigenvalue computation was measured 
according to
\[e = \frac{\Vert\vb*{\omega}_\text{ref} - \vb*{\omega}_h\Vert_2}{\Vert\vb*{\omega}_\text{ref}\Vert_2}\, ,\]
where $\vb*{u}_h$ is the set of eigenvalues (dispersion relation) for a mesh of 
characteristic element size $h$ and $\vb*{u}_\text{ref}$ is the solution 
corresponding to the $32\times32$ elements mesh, which has been taken as 
reference. The results for this sequence, together with the variation in the 
error parameter, are displayed in \cref{fig:convergence}. The estimated 
convergence rate for the eigenvalues is 1.81.

\subsection{Variation of micropolar parameters in a bilayer composite}
\begin{figure}[H]
    \centering
    \includegraphics[height=1.5 in]{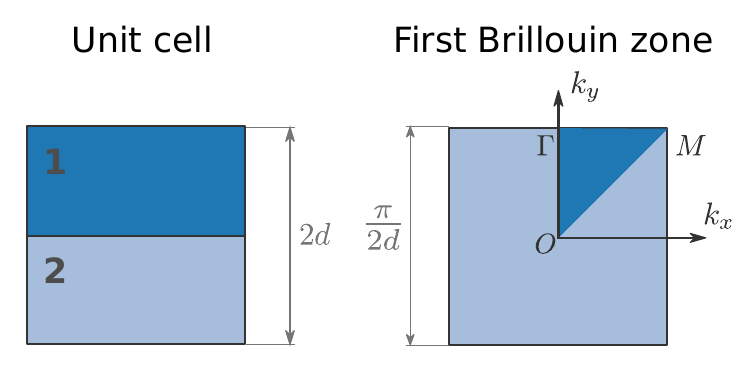}
    \caption{\small (Left) Schematic of the unit cell for the bilayer material. (Right) Illustration of the first Brillouin zone and the irreducible Brillouin zone.}
    \label{fig:bilayer_materials}
\end{figure}

We considered a bilayer composite made with two materials that share all the 
properties, except for one of the micropolar parameters. Thus, we varied $J$, 
$\alpha$ and $\xi\equiv\eta + \varepsilon$,  while keeping the other parameters 
fixed. Notice that the composite represents a homogeneous material in the case 
that both layers share the same material properties. 
\Cref{fig:bilayer_materials} shows the unit cell and the first Brillouin zone 
for this set of analyses.

The properties in material 2, as presented in \cref{fig:bilayer_materials}, are fixed. The values used are the following:
\begin{align*}
  &\rho_1 = \rho_2 = 2770 \text{ kg/m}^3\, , &\lambda_1 = \lambda_2 = 5.12\times 10^{10} \text{ Pa}\, ,\\
  &\mu_1 = \mu_2 = 2.76 \times 10^{10} \text{ Pa}\, , &\alpha_2 = 3.07\times 10^{9} \text{ Pa}\, ,\\
  &\xi_2 = 7.66 \times 10^{10} \text{ N}\, , &J_2 = 306.5 \text{ kg/m}\, .
\end{align*}

\begin{figure}[H]
    \centering
    \includegraphics[height=2 in]{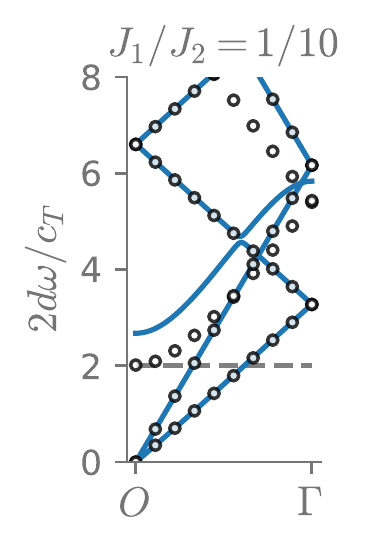}
    \includegraphics[height=2 in]{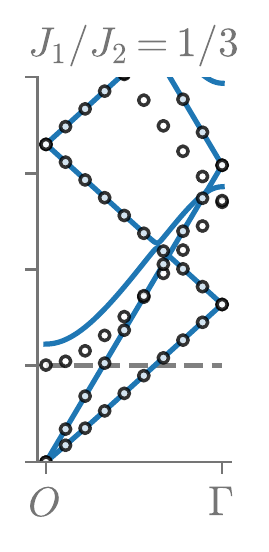}
    \includegraphics[height=2 in]{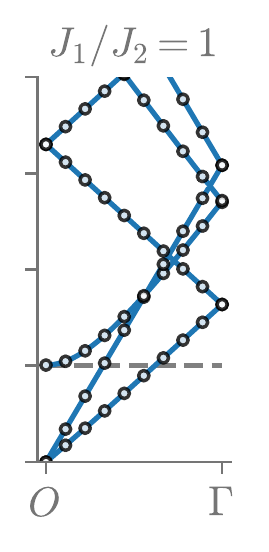}
    \includegraphics[height=2 in]{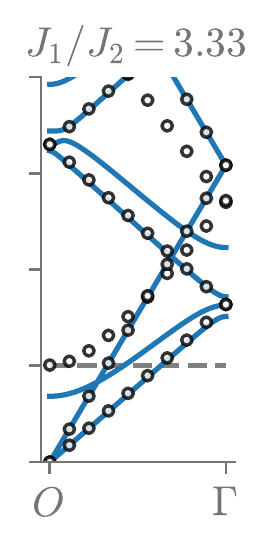}
    \includegraphics[height=2 in]{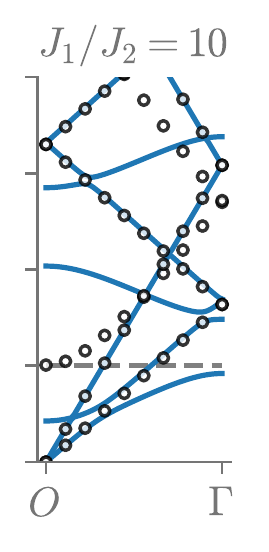}
    \caption{\small Dispersion curves for a bilayer composite made with two 
    micropolar materials. All material parameters are fixed except for the 
    inertial density $J_1$, which takes values in $\{30, 100, 300, 1000, 
    3000\}$ kg/m. The dispersion curves for the different values of $J_1$ 
    correspond to the continuous blue line, while black dots show 
    reference 
    results corresponding to a homogeneous material cell. The increase in $J_1$ 
    produces an overall increase in the inertial density accompanied by a 
    decrease in the cut-off frequency of the microrotational wave.}
    \label{fig:bilayer_var_J}
\end{figure}

\Cref{fig:bilayer_var_J} presents the results for variations in $J_1 \in \{30,\,
100,\, 300,\, 1000,\, 3000\}$ kg/m. The results are compared with those of the 
homogeneous cell shown by the black dots, while the dispersion curves resulting 
from variations in $J_1$ are described by the continuous blue line. As $J_1$ 
increases the cut-off frequency for the microrotational wave decreases, which 
is due to the overall increase in the inertial density. We can also 
highligh that for larger values of $J_1$ the dispersion for the $SV$ and $TR$ 
waves increases (see the results for $J_{1}/J_{2} = 10$). This is a result 
of the interaction between $TR$ and $SV$ waves, but $P$ waves are not affected 
since they do not interact when the incidence is perpendicular.

\begin{figure}[H]
    \centering
    \includegraphics[height=2 in]{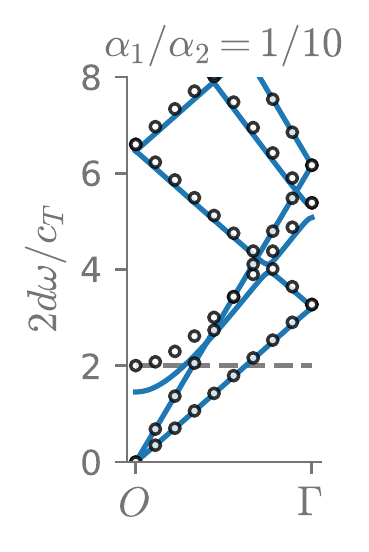}
    \includegraphics[height=2 in]{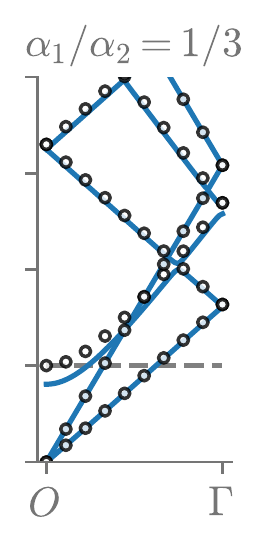}
    \includegraphics[height=2 in]{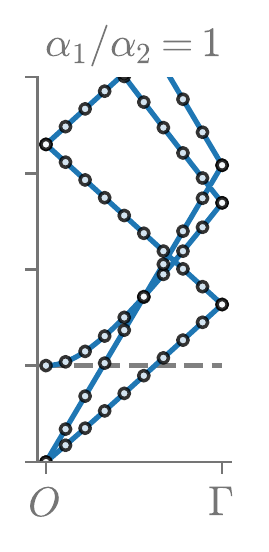}
    \includegraphics[height=2 in]{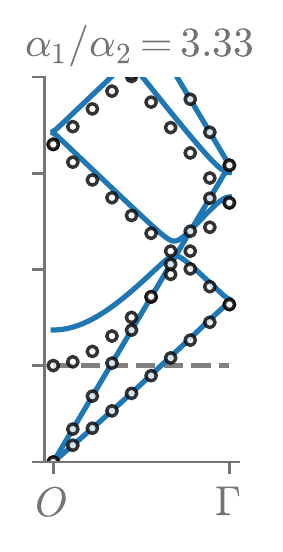}
    \includegraphics[height=2 in]{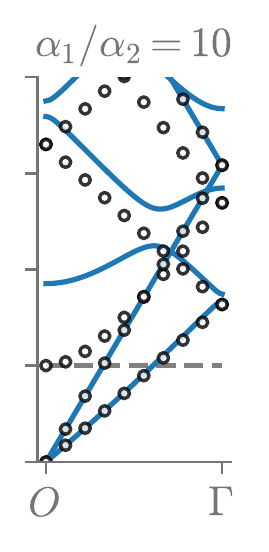}
    \caption{\small Dispersion curves for a bilayer composite made with two 
    micropolar materials. All material parameters are fixed except for 
    $\alpha_1$, which takes values in $\{3.41 \times 10^8, 1.02 \times 10^9, 
    3.07 \times 10^9, 9.21 \times 10^9, 2.76 \times 10^{10}\}$ Pa. The 
    dispersion curves for the different values of $\alpha_1$ correspond to the 
    continuous blue line, while the black dots show reference results 
    corresponding to a homogeneous material cell. The overall increase of the 
    $\alpha$ parameter for the composite produces a decrease in the cut-off 
    frequency and an increase in the dispersion for the SV waves.}
    \label{fig:bilayer_var_alpha}
\end{figure}

As a second modification to the set of material properties we now changed 
$\alpha_1 \in \{3.41 \times 10^8,\, 1.02 \times 10^9,\, 3.07 \times 10^9,\,
9.21 \times 10^9,\, 2.76 \times 10^{10}\}$ Pa. The corresponding results, in 
terms of dispersion curves are shown in \cref{fig:bilayer_var_alpha}. It is now 
evident how the cut-off frequency presents an opposite trend as compared with 
the variations in $J_1$, that is, the cut-off frequency increases as the ratio 
$\alpha_1/\alpha_2$ increases. This is also an expected result considering the 
overall increase in $\alpha$ for the composite. Furthermore, we see that we can 
increase the size of the (partial) bandgap increasing the ratio 
$\alpha_1/\alpha_2$.

We considered as a last variation changes in the parameter $\xi_1=\eta_1 + 
\epsilon_1 \in \{8.51 \times 10^8, 2.55 \times 10^9, 7.66 \times 10^9, 2.30 
\times 10^{10}, 6.89 \times 10^{10}\}$ N. Although this parameter changes the 
dispersive response, the cut-off frequency for the micro-rotational wave 
remains unmodified, as $Q$ is independent of $\xi$ and therefore the cut-off 
frequency is independent of the overall modulus of the composite. Although we 
see some (partial) bandgaps when changing the ratio $\xi_1/\xi_2$, it is more 
interesting to highlight how the hybridization between $SV$ and $TR$ modes 
changes across the different values of the ratio.
\begin{figure}[H]
    \centering
    \includegraphics[height=2 in]{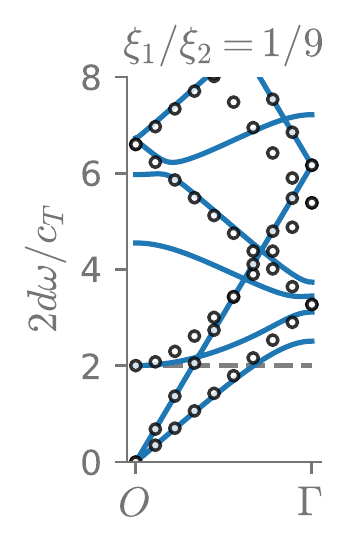}
    \includegraphics[height=2 in]{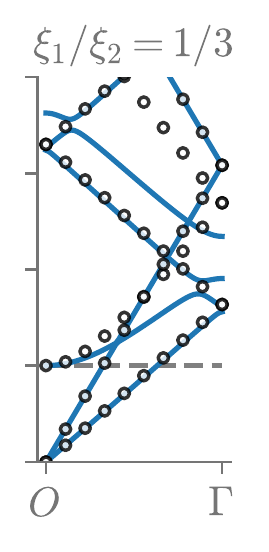}
    \includegraphics[height=2 in]{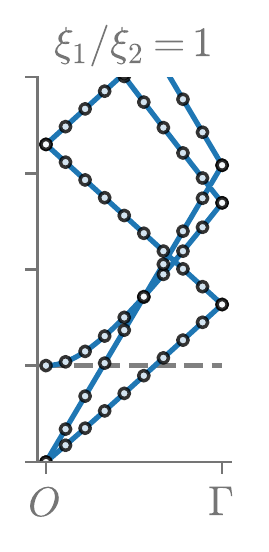}
    \includegraphics[height=2 in]{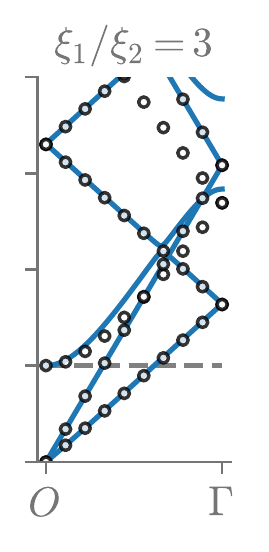}
    \includegraphics[height=2 in]{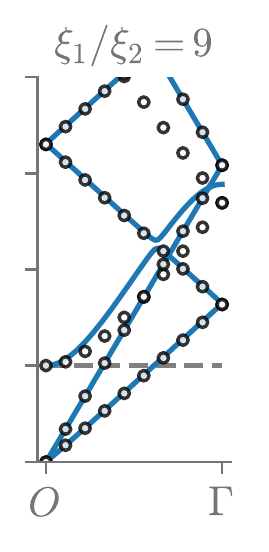}
    \caption{\small Dispersion curves for a bilayer composite made with two 
    micropolar materials. All material parameters are fixed except for $\xi_1$, 
    which takes values in $\{8.51 \times 10^8, 2.55 \times 10^9, 7.66 \times 
    10^9, 2.30 \times 10^{10}, 6.89 \times 10^{10}\}$ N. The dispersion curves 
    for the different values of $\xi_1$ correspond to the continuous blue line, 
    while the black dots show reference results corresponding to a 
    homogeneous material cell. In this case the cut-off frequency remains 
    unmodified as $Q$ is independent of $\xi$.}
    \label{fig:bilayer_var_xi}
\end{figure}

\subsection{Variation of microstructural length}\label{section:length}
\begin{figure}[H]
    \centering
    \includegraphics[height=1.5 in]{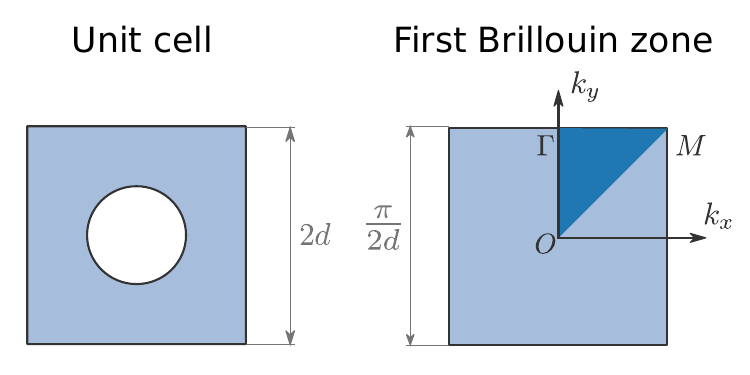}
    \caption{(Left) Schematic of the unit cell for the porous material. (Right) Illustration of the first Brillouin zone and the irreducible Brillouin zone.}
    \label{fig:pore_unit_cell}
\end{figure}

Although the micropolar medium introduces dispersive behavior through the 
presence of length scale parameters, additional frequency dependence of the 
wave propagation velocity in the medium can be obtained if we explicitly 
consider the presence of microstructural features embedded in the micropolar 
medium (\cref{fig:pore_unit_cell}). Here we explore a material cell which is 
composed of a circular pore of diameter $d$ embedded inside a micropolar medium.

We will assume a pore with a diameter that is $1/2$ of the cell length or 
equivalently a porosity of $\pi/16 \approx 0.196$. This value is kept fixed as 
we modify the values of the unit cell. We will use the following set of 
mechanical parameters for the micropolar model:
\begin{align*}
&\rho = 2770 \text{ kg/m}^3\, , &\lambda =  5.12\times 10^{10} \text{ Pa}\, ,\\
&\mu = 2.76 \times 10^{10} \text{ Pa}\, , &\alpha = 3.07\times 10^{9} \text{ Pa}\, ,\\
&\eta + \varepsilon = 7.66 \times 10^{10} \text{ N}\, , &J = 306.5 \text{ 
kg/m}\, .
\end{align*}

\begin{figure}[H]
    \centering
    \includegraphics[height=2 in]{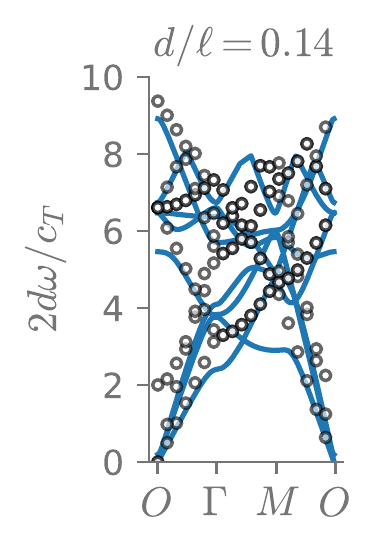}
    \includegraphics[height=2 in]{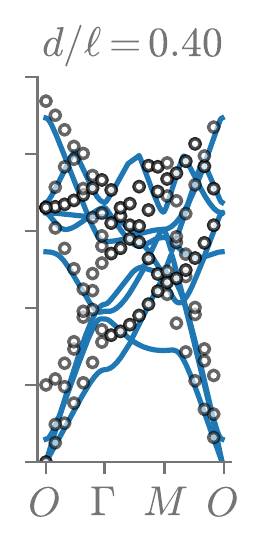}
    \includegraphics[height=2 in]{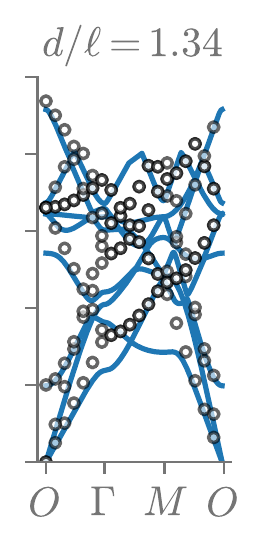}
    \includegraphics[height=2 in]{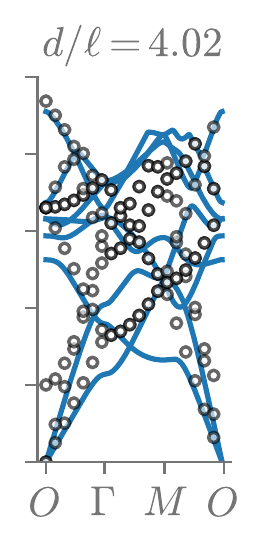}
    \includegraphics[height=2 in]{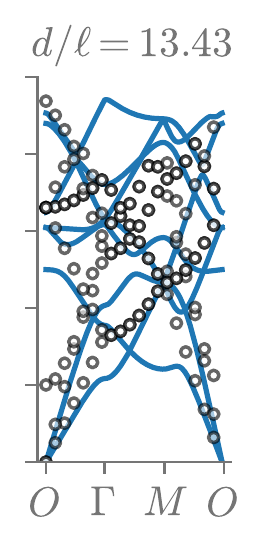}
    \caption{Variation in the band structure for a micropolar periodic 
    cellular material with circular pores for different values of the ratio 
    $d/\ell$.}
    \label{fig:circular_pore_var}
\end{figure}

For the size effect analysis, it is convenient to express the mechanical 
parameter in terms of a constitutive length scale present in the micropolar 
model as discussed in \cite{cowin1970} and given by:
\[\ell^2 \equiv \frac{\eta + \varepsilon}{2\mu} = 0.3725 \text{ m}\, ,\]
in which $2\mu \ell^2$ represents the rotational stiffness of the material 
\citep{trovalusci2015}. Notice that a change in the size of the unit cell 
implies a change in the pore diameter and therefore a change in the ratio 
between the intrinsic length scale $\ell$ and the characteristic 
microstructural dimension $d/\ell$. This variation in the dispersion relations 
with the ratio $d/\ell$ is shown in  \cref{fig:circular_pore_var}.

\subsection{Directionality in cellular material with a circular pore}
As a final result, we computed the directionality curves for cellular materials 
with circular pores. We changed the diameter of the pore keeping fixed the size 
of the cell. The material properties used are the same as in 
\Cref{section:length}. We compare the directionality results in a qualitative 
fashion, showing how different directions present different phase/group speeds 
when changing the porosity. This comparison is only valid for small wavenumber, 
because near the edges of the first Brillouin zone the different branches might 
intersect and the resulting contours would contain information from mixed modes.

\begin{figure}[H]
    \centering
    \includegraphics[height=4 in]{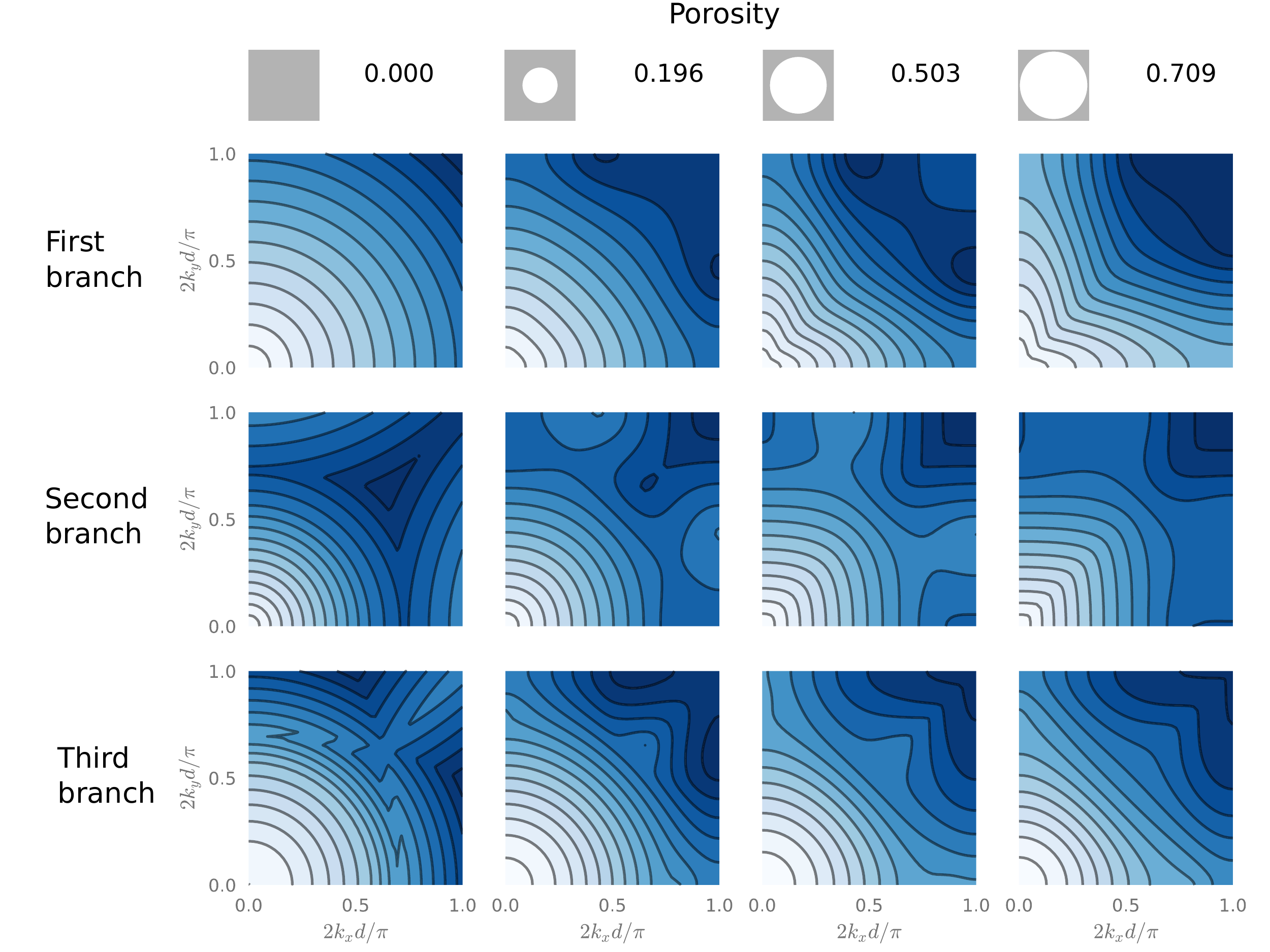}
    \caption{Directionality curves for cellular materials with increasing 
    porosity for the first three branches. The corresponding porosities are: 
    0.000, 0.196, 0.503 and 0.709. The porosity increases from left to right  
    and is presented at the top. A porosity of 0 represents a homogeneous 
    material, used as reference in this case.}
    \label{fig:direcitonality_pore}
\end{figure}

\Cref{fig:direcitonality_pore} presents the directionality curves for the first 
three branches of the dispersion relations for cellular materials with 
increasing porosity with values: 0.000, 0.196, 0.503 and 0.709. A porosity of 0 
represents a homogeneous material, used as reference in this case. The 
directionality (anisotropy) of the material increases for higher porosity 
values.

The first branch presents lower phase speed along the vertical and horizontal 
directions, while the opposite happens for the second mode. This is expected 
since these two modes represent quasi-transversal and quasi-longitudinal 
propagations modes for small wavenumbers. Furthermore, the first mode is the 
one that presents a higher change in directionality. We can also see that the 
anisotropy for the third branch is more sensitive to the wavenumber than the 
porosity, being (almost) isotropic for small wavenumbers.

\section{Conclusions}
We discussed several theoretical and simulation aspects related to a 
Cosserat-like micropolar medium. In particular, we address  the model 
capabilities to capture dispersive behavior through its kinematic assumptions 
and constitutive parameters. For that purpose, we used the theory of phononic 
crystals, where the band structure of the material is found from the analysis 
of Floquet-Bloch periodicity conditions. Within that context we considered unit 
material cells corresponding to a homogeneous material, a bilayer composite 
and a porous cell composed of a circular cavity embedded in a micropolar 
matrix. The dispersive properties in each case were measured in terms of the 
variation in the cut-off frequency of the microrotational wave for the 
different considered values of the mechanical and geometric parameters of the 
model. We found that the ability of the micropolar medium to capture 
dispersive behavior can be increased mainly through changes in the rotational 
inertia of the material and the length scale parameter associated to its 
bending stiffness. Furthermore, these results can serve as a benchmark for a 
program of physical experiments on periodic solids to explore the existence of 
microrotational waves, as predicted by micropolar theory.

\appendix
\section{Terms for the finite element equilibrium equations in the micropolar 
solid}
\label{app:dir}
The discrete finite element equilibrium equations for the micropolar solid 
were written as
\begin{equation}
\left[ \begin{array}{cc}
K_{uu}^{kp} & K_{u\theta}^{kp} \\ 
K_{\theta u}^{kp} & K_{\theta\theta}^{kp}
\end{array} \right] \left\lbrace \begin{array}{c}
u^p \\ 
\theta^p
\end{array} \right\rbrace - \omega^2\left[ \begin{array}{cc}
M_{uu}^{kp} & 0 \\ 
0 & M_{\theta\theta}^{kp}
\end{array}  \right]\left\lbrace \begin{array}{c}
u^p \\ 
\theta^p
\end{array}\right\rbrace = \left\lbrace \begin{array}{c}
f_{ext} \\ 
q_{ext}
\end{array}\right\rbrace \, ,
\end{equation}
where each one of the terms are defined next as follows,
\[K_{uu}^{kp} = \int\limits_{V} B_{ij}^k C_{ijrs}B_{rs}^{p} \dd{V} + 
\int\limits_{V} \hat{W}_{ij}^k G_{ijrs}\hat{W}_{rs}^p \dd{V} 
\equiv\int\limits_{V} 
B_{ij}^k C_{ijrs}B_{rs}^{p} \dd{V} + \int\limits_{V} \mu_c \hat{W}_{ij}^k 
\hat{W}_{ij}^p \dd{V} \, ,\]
which is symmetric.

\begin{align*}
K_{u\theta}^{kp} &= \int\limits_{V} \hat{W}_{ij}^k G_{ijrs} R_{rs}^p \dd{V} 
\equiv\int\limits_{V} \alpha \hat{W}_{ij}^k R_{rs}^p \dd{V}\\
K_{\theta u}^{kp} &= \int\limits_{V} R_{ij}^k G_{ijrs} \hat{W}_{rs}^p \dd{V} 
\equiv\int\limits_{V} \alpha R_{ij}^k \hat{W}_{rs}^p \dd{V} \, ,
\end{align*}
with $K_{u\theta}^{kp} = \left(K_{\theta u}^{kp}\right)^T$.
\[K_{\theta\theta}^{kp} = \int\limits_{V} R_{ij}^k G_{ijrs} R_{rs}^p \dd{V} + 
\int\limits_{V} M_{ij}^k D_{ijrs} M_{rs}^p \dd{V} \equiv \int\limits_{V} \alpha 
R_{ij}^k R_{rs}^p \dd{V} + \int\limits_{V} M_{ij}^k D_{ijrs} M_{rs}^p \dd{V} \, 
,\]
which is also symmetric.

Similarly, the inertial terms, which are both symmetric, are defined as
\begin{align*}
M_{uu}^{kp} &= \int\limits_{V}\rho\ _{u}  N_i^k\ _{u}N_i^p \dd{V}\\
M_{\theta\theta}^{kp} &= \int\limits_{V} J \ _{\theta}N_i^k\ _{\theta}N_j^p 
\dd{V}.
\end{align*}

Finally, the external force and couple vectors read
\begin{align*}
f_{ext}^p &= \int\limits_{S}\ _{u}N_{i}^p t_i \dd{S} \, ,\\
q_{ext}^p &= \int\limits_{S}\ _{\theta}N_{i}^p m_i \dd{S} \, .
\end{align*}

\section{In-plane equations of motion}
\label{app:2Deqs}

The equations of motion for waves in the plane are the following:
\begin{align*}
&\frac{\lambda + 2\mu}{\rho}\left[\pdv[2]{u_x}{x} + \pdv{u_y}{y}{x}\right] - \frac{\mu + \alpha}{\rho}\left[\pdv{u_y}{y}{x} - \pdv[2]{u_x}{y}\right] + \frac{2\alpha}{\rho}\pdv{\theta_z}{y} = -\omega^2 u_x\, ,\\
&\frac{\lambda + 2\mu}{\rho}\left[\pdv{u_x}{y}{x} + \pdv[2]{u_y}{y}\right] - \frac{\mu + \alpha}{\rho}\left[\pdv{u_x}{y}{x} - \pdv[2]{u_y}{x}\right] - \frac{2\alpha}{\rho}\pdv{\theta_z}{y} = -\omega^2 u_y\, ,\\
&\frac{2\alpha}{J}\left[\pdv{u_y}{x} - \pdv{u_x}{y}\right] + \frac{\eta + 
\varepsilon}{J}\left[\pdv[2]{\theta_z}{x} + \pdv[2]{\theta_z}{y}\right] - 
\frac{4\alpha}{J}\theta_z = -\omega^2 \theta_z\, ,
\end{align*}
notice that the equations do not involve the parameter $\beta$.

We can write the constitutive equations in extended Voigt's notation as
\[
\begin{Bmatrix}
\sigma_{xx} \\ \sigma_{yy}\\ \sigma_{xy}\\ \sigma_{yx}\\ \mu_{zx}\\ \mu_{zy}\\
\end{Bmatrix}
= 
\begin{bmatrix}
	\lambda + 2\mu & \lambda & 0 & 0 & 0 & 0 \\ 
	\lambda & \lambda + 2\mu & 0 & 0 & 0 & 0 \\ 
	0 & 0 & \mu + \alpha & \mu - \alpha & 0 & 0 \\ 
	0 & 0 & \mu - \alpha & \mu + \alpha & 0 & 0 \\ 
	0 & 0 & 0 & 0 & \eta + \varepsilon & 0 \\ 
	0 & 0 & 0 & 0 & 0 & \eta + \varepsilon
\end{bmatrix}
\begin{Bmatrix}
\gamma_{xx} \\ \gamma_{yy}\\ \gamma_{xy}\\ \gamma_{yx}\\ \kappa_{zx}\\ \kappa_{zy}\\
\end{Bmatrix}\, ,\]
where
\begin{equation*}
\begin{split}
\gamma_{xx} = \pdv{u_x}{x},\quad &\gamma_{yy} = \pdv{u_y}{y},\\
\gamma_{xy} = \pdv{u_y}{x} + \theta_z,\quad &\gamma_{yx} = \pdv{u_x}{y} - \theta_z,\\
\kappa_{zx} = \pdv{\theta_z}{x},\quad &\kappa_{zy} = \pdv{\theta_z}{y}.\\
\end{split}
\end{equation*}

\section*{Acknowledgements}
This work was supported by EAFIT and COLCIENCIAS' Scholarship Program No. 6172.

\section*{References}
\bibliographystyle{unsrtnat}
\bibliography{bloch-cosserat}

\end{document}